\documentclass{article}  
\usepackage{spadre2008}
\usepackage{graphicx}
\frompage{000} \topage{000}                                              
\def\rightmark{Charmonium Production at High $p_T$ at RHIC}
\def\leftmark{X.Zhao and R.Rapp}
 
\title{Charmonium Production at High $p_t$ at RHIC} 
\authors{
{Xingbo Zhao and Ralf Rapp %
}\\[2.812mm]
{\normalsize
\hspace*{-8pt} Cyclotron Institute and Physics Department, 
Texas A\&M University, \\ 
77843-3366 College Station, USA\\[0.2ex] 
}}
 
\abstract{We calculate transverse-momentum ($p_t$) spectra and 
  elliptic flow of $J/\psi$'s in $Au$-$Au$ and $Cu$-$Cu$
  collisions at RHIC. We employ an earlier constructed 2-component 
  approach for direct and regenerated charmonia which is fairly  
  consistent with the centrality dependence of inclusive $J/\psi$ 
  yields at SPS and RHIC. Discrepancies with high-$p_t$ data 
  in $Cu$-$Cu$ collisions are addressed by implementing effects of 
  a finite formation time and an estimate of $B$ meson feeddown,
  which largely resolve the problem. The sensitivity of the spectra 
  to the Cronin effect and different in-medium dissociation rates
  is studied. The predicted elliptic flow in semicentral $Au$-$Au$ 
  collisions is rather small, $\sim$2-3\%.}

 
\begin{document}
 
\maketitle
\setcounter{page}{1}

\section{Introduction}\label{intro}

An ``anomalous"  suppression of $J/\psi$ mesons produced in ultrarelativistic
heavy-ion collisions (URHICs) has long been suggested as a signature
of Quark-Gluon Plasma (QGP) formation~\cite{Matsui:1986dk}. However, 
experiments at the CERN Super Proton Synchrotron (SPS) and the 
Brookhaven Relativistic Heavy-Ion Collider (RHIC) show a very comparable 
suppression of $J/\psi$'s, despite much larger energy densities believed 
to be created at RHIC compared to SPS. Such a result has, in fact, been 
predicted in approaches~\cite{Grandchamp:2001pf} where suppression 
mechanisms are supplemented by secondary charmonium production, e.g., 
via the coalescence of $c$ and $\bar{c}$ quarks at hadronization
transition~\cite{BraunMunzinger:2000px,Gorenstein:2000ck}.
However, significant uncertainties in these approaches remain, most
notably due to the total charm cross section and the degree of 
thermalization of the charm-quark spectra. It is therefore essential 
to find additional means to discriminate suppression and regeneration
mechanisms. Transverse-momentum 
($p_t$)~\cite{Bugaev:2002xx,Andronic:2006ky,Greco:2003vf,Thews:2005vj,Yan:2006ve} 
and rapidity ($y$)~\cite{Andronic:2006ky} spectra, as well as elliptic 
flow~\cite{Greco:2003vf,Yan:2006ve,Wang:2002ck,Ravagli:2007xx} of 
$J/\psi$'s are hoped to provide such tools. 

In the present paper we first discuss our recent 
extension~\cite{Zhao:2007hh} of the 
two-component model~\cite{Grandchamp:2001pf,Grandchamp:2003uw}
to compute $p_t$ spectra, augmented by predictions for elliptic 
flow, $v_2(p_t)$, in Sec.~\ref{sec_2comp}. In Sec.~\ref{sec_hipt} we 
evaluate additional mechanisms relevant to $J/\psi$ production at high 
$p_t>$5GeV/$c$. Sec.~\ref{sec_concl} contains our conclusions.
 
\section{$p_t$ Spectra and Elliptic Flow in the 2-Component Approach}
\label{sec_2comp}  
\begin{figure} [!t]
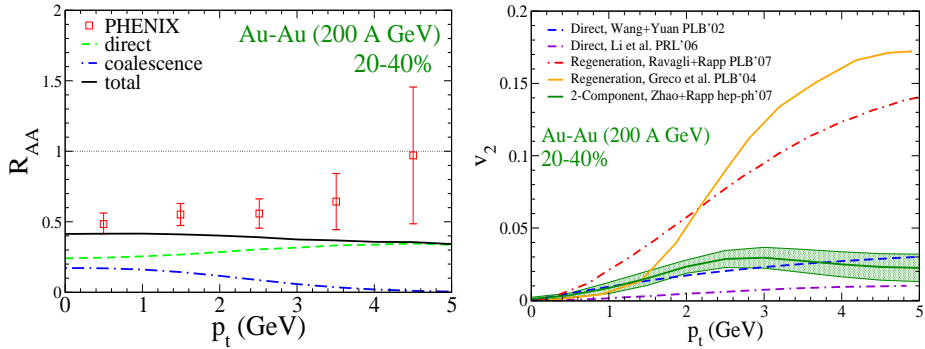

\includegraphics[width=0.47\textwidth,height=0.36\textwidth]{raapt7.8_v2.eps}
\hspace{0.0cm}
\includegraphics[width=0.47\textwidth,height=0.36\textwidth]{v2_b7.8.eps}
\caption{(Color online). Left panel: 2-component approach for $J/\psi$
  $R_{AA}(p_t)$ in 20-40\% $Au$-$Au$ at RHIC, compared to PHENIX 
 data~\cite{Adare:2006ns}. Right panel: estimated $v_2(p_t)$
  based on the partition in the left panel. The band represents
  uncertainties due to differences between two
  independent input $v_2(p_t)$ calculations for each of the two
  components.}
\label{fig_rhic2040}
\end{figure}

Within the two-component approach, the $p_t$ spectra of charmonia 
($\Psi=J/\psi, \chi_c, \psi'$) in URHICs are decomposed into 
contributions from direct production and 
regeneration~\cite{Zhao:2007hh}. The former 
is based on initial spectra in elementary $N$-$N$ collisions including 
a Cronin effect (i.e., nuclear $p_t$ broadening implemented via a 
Gaussian smearing) and primordial nuclear absorption estimated from
$p$-A collisions; the resulting spectra serve as an input to a Boltzmann 
transport equation to compute subsequent suppression in an expanding 
thermal fireball through the QGP, mixed and hadron gas (HG) phase. 
The charmonium dissociation
rates in the QGP account for in-medium reduced binding energies and
are therefore calculated using inelastic quasifree cross 
sections~\cite{Grandchamp:2001pf} (rather than gluo-dissociation~\cite{Peskin:1979va}),
while HG dissociation is estimated using $SU$($N_f$=4) effective theory.
We also take into account the ``leakage" 
effect~\cite{Karsch:1987,Blaizot:1988ec}, i.e.,  charmonia traveling 
outside the fireball boundary before freeze-out are not subject to 
dissociation. The leakage effect reduces suppression primarily
for high-$p_t$ charmonia.
The inclusive yield of the regeneration component is computed from 
$c$-$\bar c$ coalescence using detailed balance (i.e., the inverse of 
quasifree dissociation) within a momentum-independent thermal rate
equation~\cite{Grandchamp:2003uw}. Regeneration processes are 
suppressed by a schematic relaxation-time factor to simulate 
incomplete kinetic equilibration of the charm quarks. 
The $p_t$ spectra of regenerated $\Psi$'s are approximated 
by a local thermal blastwave distribution~\cite{Schnedermann:1993ws} 
at the hadronization transition
based on the same fireball as used for the direct component.    
The input charm cross section, $\sigma_{\bar cc}$=570$\mu$b, 
is consistent with recent PHENIX data~\cite{Adler:2004ta}. 
Our approach yields a reasonable overall description of the
nuclear modification factor, 
$R_{AA}(N_{part})$, for inclusive $J/\psi$ production, as well
as its $p_t$ dependence in $Pb$-$Pb$ and $Au$-$Au$ collisions
at SPS and RHIC, respectively~\cite{Zhao:2007hh}. At SPS and
for peripheral collisions at RHIC, the regeneration contribution
is small, but it becomes significant for $p_t\le$2-3~GeV for 
semi-/central collisions $Au$-$Au$ collisions at RHIC, see, e.g.,
the left panel of Fig.~\ref{fig_rhic2040}.

We now apply our approach to estimate the elliptic flow of $J/\psi$'s 
according to
\begin{equation}
v_2(p_t) = f^{dir} v_2^{dir}(p_t) + f^{reg} v_2^{reg}(p_t) \ ,
\end{equation}
where the fractions $f^{dir}$ and $f^{reg}$ follow from the 
decomposition provided by the two-component model for 20-40\%
central $Au$-$Au$ at RHIC. For the underlying $v_2$ spectra
we take guidance from the suppression
calculations in Refs.~\cite{Yan:2006ve,Wang:2002ck} for the direct
component, $v_2^{dir}(p_t)$, and from the coalescence models
in Refs.~\cite{Greco:2003vf,Ravagli:2007xx} for the regenerated
component, cf.~the right panel in Fig.~\ref{fig_rhic2040}. The
weighted sum is shown by the green band reflecting the uncertainties
in the input $v_2$ of the 2 components. It is somewhat sobering to
find that the resulting $v_2(p_t)$ is small, around 2-3\%, over the 
entire $p_t$ range. The reason is that a large $v_2$ for the coalescence
component only develops above $p_t$$\simeq$2~GeV, where, however, its
weight is already small. On the other hand, in the low $p_t$ region,
where the coalescence component carries a significant weight, 
its $v_2$ is still small. The $v_2$ of the direct component, which solely
derives from the path length dependence relative to the reaction plane,
is always small. This puts a rather high requirement on the
accuracy of future measurements of $J/\psi$ elliptic flow in order
to extract information about $J/\psi$ production mechanisms.

\section{Charmonium Production at High $p_t$}
\label{sec_hipt}
\begin{figure}[!t]
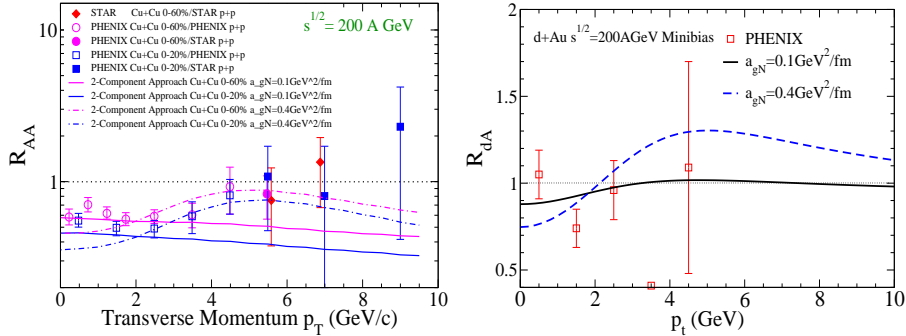

\begin{center}
\includegraphics[width=0.46\textwidth,height=0.35\textwidth]{raa_pt_agn0.4_cu.eps}
\hspace{0.0cm}
\includegraphics[width=0.46\textwidth,height=0.35\textwidth]{raa_mb_pt_dAu.eps}
\end{center}
\caption{(Color online). 
Left panel: $J/\psi$ $R_{AA}(p_t)$ in $Cu$-$Cu$ at RHIC within  our 
  2-component approach using $a_{gN}$=0.1,0.4~GeV$^2$/fm (solid, dot-dashed 
  lines), compared to PHENIX~\cite{Adare:2008sh}
  and STAR data~\cite{Tang:2008uy}. 
Right panel: Cronin effect for $J/\psi$ $R_{dA}(p_t)$ 
  implemented via Gaussian smearing with
  $a_{gN}$=0.1,0.4~GeV$^2$/fm (solid, dashed line), 
   compared to PHENIX $d$-$Au$ data~\cite{Adare:2007gn}.} 
\label{fig_cronin}
\end{figure}

The predictions of our approach for $p_t$ spectra in 200~AGeV 
$Cu$-$Cu$ collisions are compared to data in the left panel 
of Fig.~\ref{fig_cronin} (solid lines).
While the agreement is fair up to $p_t\simeq 4$~GeV, at 
higher momenta a significant discrepancy develops.  
In the following, we will investigate several potential sources of this 
discrepancy related to high-$p_t$ production which have not been 
included in our baseline calculations~\cite{Zhao:2007hh}. 

First we check whether the increasing trend of $R_{AA}(p_t)$ can be 
attributed to a stronger Cronin effect, which at RHIC energies is 
currently not well constrained by $d$-$Au$ data, cf.~the right panel of
Fig.~\ref{fig_cronin}. In our baseline calculations, we used the same
broadening as at SPS, according to $\langle p_t^2\rangle_{AA} = \langle
p_t^2\rangle_{pp} + a_{gN} L$ with $a_{gN}$=0.1GeV$^2$/fm  ($L$ is
an average gluon path length prior to the hard process for charm
production). With a larger $a_{gN}$=0.4GeV$^2$/fm, which might
still be acceptable for the $d$-$Au$ data, the high $p_t$ regime
in the $Cu$-$Cu$ spectra indeed improves (dashed lines in the right panel
of Fig.~\ref{fig_cronin}), which is, however, somewhat 
at the expense of the low-$p_t$ region. Clearly, more
accurate $d$-$Au$ data will be very valuable to constrain the 
initial nuclear $p_t$ broadening. 

\begin{figure} [!t]
\includegraphics[width=0.47\textwidth,height=0.35\textwidth]{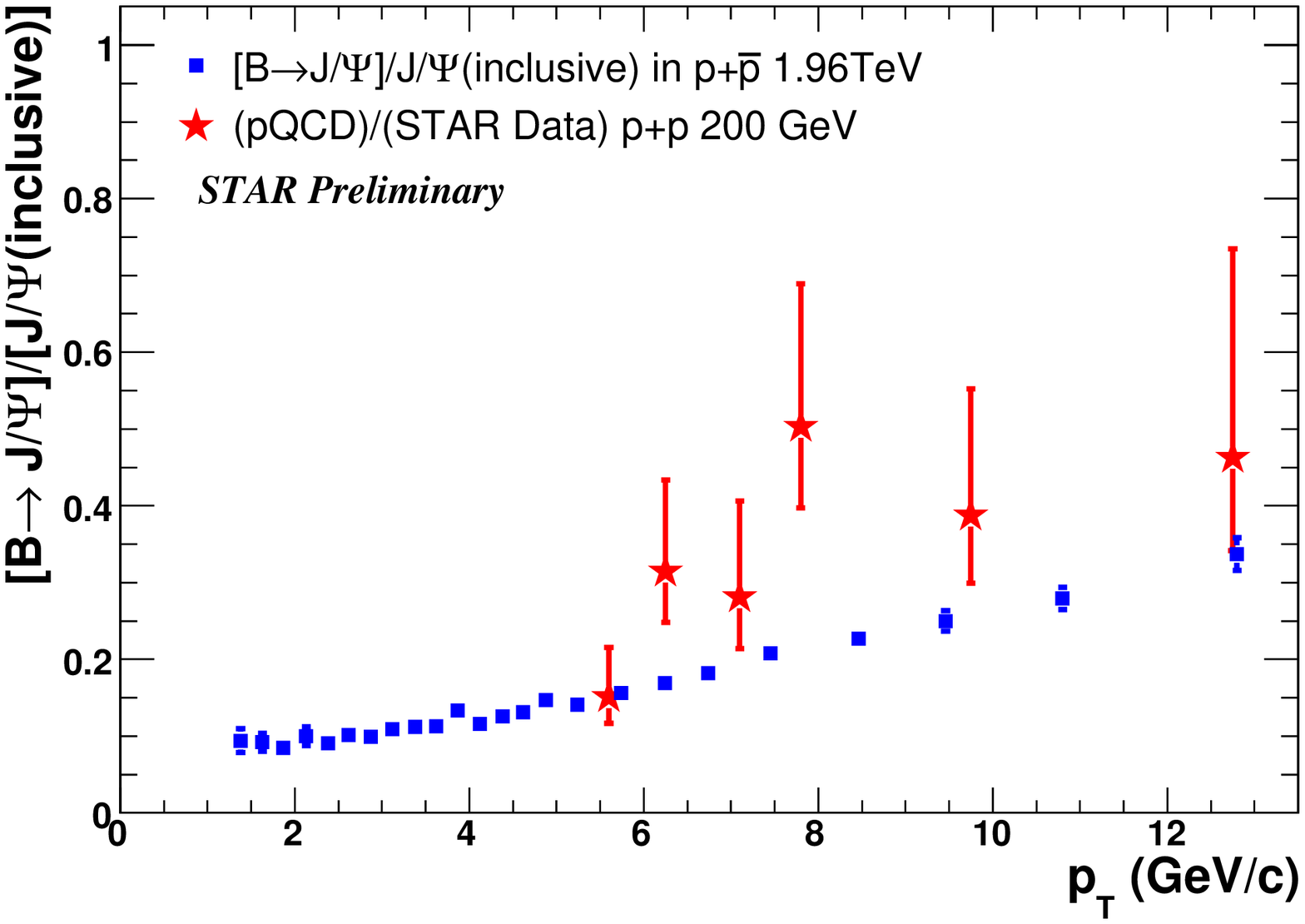}
\hspace{0.0cm}
\includegraphics[width=0.47\textwidth,height=0.35\textwidth]{raa_pt_cu_formtime.eps}
\caption{(Color online). Left panel: Tevatron~\cite{Acosta:2004yw} and 
  STAR data for 
  the ratio of $J/\psi$ from $B$-meson feeddown to inclusive 
  $J/\psi$~\cite{Tang:2008private}. Right panel: $J/\psi$
  $R_{AA}(p_t)$ within the 2-component approach including formation 
  time effects and $B$-meson feeddown (green band) in 0-60\% $Cu$-$Cu$,
  compared to our previous results (solid lines) and RHIC 
  data~\cite{Adare:2008sh,Tang:2008uy}.}
\label{fig_form}
\end{figure}

Second, we consider the following two effects primarily relevant
at high $p_t$: 
(1) Finite formation times~\cite{Blaizot:1988ec,Gavin:1990gm,Karsch:1988}
and (2) Bottom feeddown. 
Concerning (1), one expects reduced dissociation cross
sections for a ``pre-hadronic" $c\bar{c}$ pair relative to
a fully formed charmonium due to a finite formation time, $\tau_{f}$,
required to build up the hadronic wave function. For a schematic estimate 
we parameterize the evolution of the pre-hadronic dissociation rate as
\begin{equation}
\Gamma_{c\bar c}=\Gamma_{\Psi} \tau/\tau_{f}^{lab} 
 \quad , \quad \tau \le \tau_{f}^{lab} = \tau_{f} m_t/m_\Psi  
\label{form_time}
\end{equation}
with $\Gamma_{\Psi}$: (nuclear, partonic or hadronic)
dissociation rates for a formed charmonium, $\tau$: fireball proper 
time, $\tau_{f}$=0.89(2.01,1.50)~fm/$c$: formation time 
of $J/\psi$($\chi_c$, $\psi^{\prime}$) in its rest frame,
and $m_t=(m_\Psi^2+p_t^2)^{1/2}$. In essence, (pre-) charmonium 
dissociation rates acquire an additional momentum dependence 
through Lorentz time dilation, being reduced at high $p_t$. 
Also note that the longer formation times of $\chi_c$ and
$\psi^{\prime}$ imply less suppression relative to $J/\psi$, 
quite contrary to standard dissociation and statistical 
regeneration mechanisms, thus constituting a rather unique signature
of the formation time effect. 
Concerning (2), the left panel of Fig.~\ref{fig_form} shows recent 
data on the $B\to J/\psi$ feeddown fraction in elementary 
$p$-$p$($\bar p$) collisions, which is quite significant. 
As an estimate of this contribution, we use the Tevatron 
data~\cite{Acosta:2004yw} with an extra uncertainty of $\pm$50\%
and replace that part of our direct component. 
Combining formation time effects and $B$-meson feeddown we obtain
as our current best estimate the green band in the right panel of 
Fig.~\ref{fig_form} which is not far off the experimental data. 

\begin{figure} [!h]
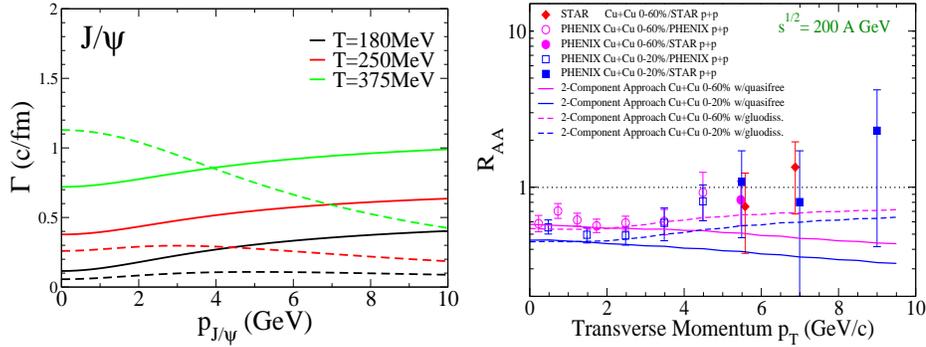

\includegraphics[width=0.47\textwidth,height=0.36\textwidth]{psi.eps}
\hspace{0.0cm}
\includegraphics[width=0.47\textwidth,height=0.36\textwidth]{raa_pt_gluodiss_cu.eps}
\caption{(Color online). Left panel: Comparison~\cite{Zhao:2007hh} of 
  the momentum
  dependence of quasifree (solid lines) and gluo-dissociation rates
  (dashed lines) for $J/\psi$ at different temperatures. Right panel:
  $J/\psi$ $R_{AA}(p_t)$ with gluo-dissociation (dashed
  lines) compared to quasifree destruction (solid lines) and
  RHIC data~\cite{Adare:2008sh,Tang:2008uy}. 
  Formation time effects and $B$ meson feeddown are not included.}
\label{fig_gluodiss}
\end{figure}

Finally, we compare the impact of different QGP dissociation mechanisms 
on $J/\psi$ $p_t$ spectra. In addition to quasifree destruction, 
we have employed the traditional gluo-dissociation~\cite{Peskin:1979va} 
(using vacuum binding energies), which exhibits a significantly
softer 3-momentum dependence, cf.~left panel of Fig.~\ref{fig_gluodiss}.
This indeed leads to somewhat harder $p_t$ spectra (see right
panel of Fig.~\ref{fig_gluodiss}), but we note that the use of 
vacuum $J/\psi$ binding energies in the QGP may not be very realistic.

\section{Conclusions}
\label{sec_concl}
Employing a 2-component model extended to calculate charmonium $p_t$ 
spectra in URHICs, we have computed $J/\psi$ elliptic flow and studied
additional effects relevant at high $p_t$. We find a rather small 
$v_2$ due to the smallness of the regeneration component 
at high $p_t$. A fair description of available high-$p_t$
spectra seems to be possible upon inclusion of formation time effects 
and $J/\psi$'s from $B$-meson feeddown. 

\section*{Acknowledgments}
We thank the workshop organizers for a very informative meeting, and 
L. Grandchamp and H. van Hees for numerous helpful discussions. This work 
is supported by a US National Science Foundation CAREER award under grant
No. PHY-0449489.


\vfill\eject
\end{document}